\documentclass[12pt,a4paper]{article}

\usepackage{amsfonts}
\newcommand{\journal}[1]{{#1}}
\newcommand{\vol}[1]{{\bf #1}}

\newcommand{\ttitle}[1]{{\it #1}}
\newcommand{\tpretitle}[1]{}

\newcommand{\inproctitle}[1]{``#1''}

\newcommand{\tref}[1]{(\ref{#1})}

\newcommand{\tpre}[1]{}
\newcommand{\tprenote}[1]{}

\newcommand{\tnote}[1]{}
\newcommand{\tcomment}[1]{}

\newcommand{\href}[2]{{#2}}
\newcommand{\eprint}[1]{{\tt #1}}

\newcommand{\tsedevelop}[1]{{}}

\makeatletter

\@addtoreset{equation}{section}
\makeatother
\newcommand{\nnel}{\nonumber \\ {}}

\newcommand{\half}{{\frac{1}{2}}}

\newcommand{\Fft}{{e^{2i  (t-t_i)  \Delta_0 z} }}
\newcommand{\Ffmt}{{e^{- 2i  (t-t_i)  \Delta_0 z} }}

\newcommand{\Nf}{{N_f}}
\newcommand{\dtf}{{i (t-t_i) }}
\newcommand{\dtfob}{\frac{i(t-t_i)}{\beta}}

\newcommand{\bfk}{{\veck}}
\newcommand{\bfp}{{\vecp}}

\newcommand{\bfx}{{\vecx}}

\newcommand{\bfkpp}{{{\veck}+{\vecp}/2 }}
\newcommand{\bfkmp}{{{\veck}-{\vecp}/2 }}
\newcommand{\epsk}{{\epsilon_{\veck}}}

\newcommand{\delo}{{\Delta_0}}
\newcommand{\Ek}{{E_\bfk}}

\newcommand{\Ekpp}{{E_\bfkpp}}
\newcommand{\Ekmp}{{E_\bfkmp}}

\newcommand{\paa}{{\partial}}

\newcommand{\Edd}{E''}

\newcommand{\Seff}{S_{\rm eff}}
\newcommand{\Sefft}{S_{\rm eff}^{(2)}}
\newcommand{\Seffbar}{\bar{S}_{\rm eff}}

\newcommand{\veck}{{\vec{k}}}
\newcommand{\vecp}{\vec{p}}
\newcommand{\vecx}{\vec{x}}
\newcommand{\vecy}{\vec{y}}
\newcommand{\vecnab}{\vec{\nabla}}

\newcommand{\psivec}{\mathbf{\Psi}}

\newcommand{\intdtk}{\int \frac{d^3\bfk}{(2\pi)^3}  \;}
\newcommand{\intdtp}{\int \frac{d^3\bfp}{(2\pi)^3}  \;}

\newcommand{\intdfx}{\int d^4x \;}

\newcommand{\calB}{{\cal B}}

\newcommand{\Abar}{\bar{A}}
\newcommand{\Zbar}{\bar{Z}}

\newcommand{\calD}{{\cal{D}}}
\newcommand{\nn}{\nonumber}

\newcommand{\matr}[1]{\mathbf{#1}}

\newcommand{\matG}{\matr{G}}
\newcommand{\matK}{\matr{K}}
\newcommand{\matphi}{\matr{\Phi}}
\newcommand{\matone}{\mathrm{\hbox{{\bf 1}\kern-.26em{\bf I}}}}

\newcommand{\tsespace}{\; \; \;  \; \; \; \; \; \; \; \;}

\begin{document}

\typeout{--- Title page start ---}

\renewcommand{\thefootnote}{\fnsymbol{footnote}}

\begin{flushright}
Imperial/TP/98-99/50
\\ DAMTP-1999-61
%\\ {\tt cond-mat/9909103}
%\\ \tsecompldate 
\end{flushright}

\vskip 12pt

\begin{center}
{\large\bf Non-equilibrium, time-dependent effective theory
of a weakly coupled
superconductor at finite temperature.}\\
\vskip 1.2cm
{\large
T.S.Evans$^a$\footnote{E-mail:
\href{mailto:T.Evans@ic.ac.uk}{\tt T.Evans@ic.ac.uk},
WWW: \href{http://euclid.tp.ph.ic.ac.uk/links/time}{\tt
http://euclid.tp.ph.ic.ac.uk/\symbol{126}time}
}
and  D.A.Steer$^b$\footnote{E-mail:
\href{mailto:D.A.Steer@damtp.cam.ac.uk}{\tt D.A.Steer@damtp.cam.ac.uk}}
}\\
\vskip 5pt
{\it a})
\href{http://euclid.tp.ph.ic.ac.uk/}{Theoretical Physics},
 Blackett Laboratory, Prince Consort Road, Imperial College,
London, SW7 2BZ, U.K.
\vskip 3pt
{\it b}) D.A.M.T.P., Silver Street, Cambridge, CB3 9EW, U.K.\\

\end{center}

%\begin{center}
%Tel: +44-1223-338957 \\
%Fax: +44-1223-337918 \\
%\mbox{  }\\
%PACS: 11.10.Wx, 11.10.Kk
%\\
%\mbox{ } \\
%{\bf Keywords} \\
%Finite temperature, chemical potential,2 dimensions
%\end{center}

%See pp.175
\renewcommand{\thefootnote}{\arabic{footnote}}
\setcounter{footnote}{0}
\typeout{--- Main Text Start ---}

\vskip 1cm
\begin{abstract}

We perform a well defined derivative expansion to obtain the time dependent
effective theory for a BCS superconductor at finite temperature, using
an arbitrary curve in the complex time plane.
Our expansion is unique, being free of any problems related to the
order in which analytical continuation and the derivative
expansion are made, or equivalently to the order in which the zero
momentum and zero energy limits are taken.  In other words Landau
damping terms are present but do not lead to the pathologies found
in standard approaches.  We discuss the reason for this,
suggesting that these standard approaches may be internally
inconsistent.  The methods presented here are an attempt to
overcome those inconsistencies, and finally, our results are
compared to others in the literature.

\end{abstract}

\vspace{1cm}

% ****************************************************************

\section{Introduction}\label{intro}

Effective actions have many applications in physics, since they
simplify the description of highly complicated physical systems by
integrating out all but the relevant degrees of freedom.  As a
result one is left with effective equations for the physically
significant modes, with coefficients and couplings depending on the
rest of the content of the theory.   In the case of superconductors,
static phenomena are well described by the effective Lagrangian of
Ginzburg and Landau \cite{GL}, and this can be derived from the
microscopic BCS theory (see for example \cite{Popov,AdriaanLH} and
references within).  In relativistic particle physics,
effective actions play an important r\^ole in inflationary theories of
cosmology \cite{KT}, and can be used to describe low energy QCD
and electro-weak phenomena \cite{DGMP} to name but a few.  In the
simplest case of fields which are constant
in space and time, the effective action reduces to the
effective potential multiplied by a space-time volume
factor.

However, to study the non-equilibrium dynamics say of a phase
transition, be it in superconductors, inflation, or QCD, space and
time dependent corrections to the effective potential are
required.  To obtain these, one often carries out a derivative
expansion, the lowest order of which gives the effective potential.
Calculations of the derivative expansion have been successfully
carried out at zero temperature $T=0$ where there is no
Landau-damping, e.g.\ see \cite{AdriaanLH,Adr,AitchSeattle} in the
case of BCS superconductors, and references
\cite{DGMP,Co,Fr} for relativistic
contexts. But at finite temperature, $T>0$, the problems arising
from Landau-damping have led to a failure of the derivative
expansion both for BCS theory \cite{AT,St,vW,AL} and in
relativistic models \cite{Fu2,TSEze,TSEwpg,EEV,GH,We,AVBD,Vo,DH}.

So far, suggested solutions to this problem have been limited.
One common approximation which is often
used to produce an effective theory is simply to replace the classical
potential by the finite temperature effective potential,
keeping the same classical kinetic terms.  Indeed this approach is often used
in simple models of inflation \cite{KT} and bubble nucleation in first
order phase transitions (see for example \cite{Anne}).
It is very difficult to justify
such methods, however, as the derivative expansions which should be
used to verify them fail at $T>0$ as commented above.
In the case of BCS superconductors, Stoof \cite{St}
appreciated the problems but simply dropped the Landau damping terms.
The same is true of \cite{Popov}.
Aitchison and Lee \cite{AL} try another approach.  They include
the effect of thermal damping, i.e.\ resum some effects from
higher order terms into the propagators \cite{AL}.  This solves
the technical problem with the derivative expansion and is
physically realistic, but experience
suggests that it can be difficult to implement such
resummations consistently \cite{Ko,KRS}. It is also technically
very difficult to work with these methods.  Many other authors have
also tackled this question (see \cite{AdriaanLH} and references
within).

The purpose of this paper is to study once more the question of
derivative expansions at $T>0$, and here it is done in the context
of BCS theory.  As mentioned above, this question has been the subject
of numerous papers, not only in the case of superconductors
\cite{AdriaanLH,AT,St,vW,AL}, but also in the application of derivative
expansions to particle physics and cosmology
\cite{Fu2,TSEze,TSEwpg,EEV,GH,We,AVBD,Vo,DH}.  Despite this, we feel
that this paper is justified as it raises some questions and potential
problems with those previous calculations.  Indeed, we will argue that
we believe those calculations to be internally  inconsistent and
therefore incorrect.  It is then argued that this inconsistency is
responsible for the badly defined and non-unique $T>0$
derivative expansions which were previously obtained.

The paper is set up in the following way.  In section \ref{secea}
the effective action for the pair field is derived from the BCS
Lagrangian.  This effective action can only be evaluated within an
approximation scheme, whose first component is a weak field
expansion (section \ref{secwf}).  The second component is a
derivative expansion (sections \ref{secde}-\ref{secres}).  Section
\ref{secde} discusses the reasons for which we believe the
previous calculations of the derivative expansion to be internally
inconsistent, whilst section \ref{seccalc} presents our own
calculation scheme.  Throughout we work at finite temperature and
in {\em arbitrary} path ordered approaches to thermal field
theory---that is, we do not work specifically in the imaginary- or
real-time formalism, but with an arbitrary contour $C$ in the
complex time plane.  As usual, the contour starts at $t_i$ and
ends at $t_i - i \beta$ (where $\beta = 1/T$), and our results are
explicitly shown to be the same for any $C$.  To carry out the
calculation we work throughout in $(t,\bfp)$ space, and
furthermore on its implementation in section \ref{seccalc} we see
that Landau-damping no longer leads to problems with our
derivative expansion which turns out to be well defined.  The
explicit results of our calculation are given in section
\ref{secres}.  There we work for simplicity in the degenerate gas
approximation \cite{Popov} and present results for the effective
potential as well as higher order corrections in time and space
gradients.  From those results the effective equations of motion
for the Cooper pair field are seen to {\em break} time translation
invariance, having an {\em explicit} dependence on the initial
time $t_i$ at which the system was set up.  We believe this to be
reasonable as by the very act of performing a derivative expansion
one is  dealing with time-dependent non-equilibrium systems. Our
effective equations are also seen to contain both oscillatory and
dissipative terms for all times $t > t_i$. Finally, though we
believe that the method presented here is an improvement on
previous ones, it raises some confusing issues and may still not
be entirely satisfactory.  Such issues are discussed with the
conclusions in section \ref{secconc}.

% ****************************************************************
\section{The effective action}\label{secea}

In this section we derive the effective action for the auxiliary
(pair) field $\Delta$ \cite{Popov,AdriaanLH}.  We work in Minkowski
space in (3+1) dimensions and use the notation $x = ( t ,\vecx
)$ where, as usual in finite temperature calculations, the
times can be complex.
The starting point is the microscopic BCS
Lagrangian
\begin{equation}
{\cal{L}} =
\sum_{s=\uparrow,\downarrow} \psi^{*}_{s} \left( i \paa_t +
\frac{\nabla^2}{2m} + \mu \right) \psi_{s} +
g \psi^{*}_{\uparrow}
\psi^{*}_{\downarrow} \psi_{\downarrow} \psi_{\uparrow}.
\label{LBCS}
\end{equation}
Here $\psi_{s}$ are anticommuting fermionic field operators, $m$ is
the electron mass and $\mu$ is the chemical potential.
The last term leads
to the formation of bound states (Cooper pairs) for $g > 0$.
As usual,
it is useful to work with the two
component fields
$$
\psivec = \left( \begin{array}{c} \psi_{\uparrow} \\
           \psi_{\downarrow}^*  \end{array} \right);
     \:\:\:\:\:\:
     \psivec^{\dagger} = (\psi_{\uparrow}^{\ast},\psi_{\downarrow})
$$
so that in equilibrium at finite temperature $T=1/\beta$ the partition
function is
\begin{equation}
Z = \int \calD \psivec \calD \psivec^{\dagger} \exp \left[ i  \int_C
d^4 x
 {\cal{L}} \right].
\label{partZ} \end{equation}
Here $\int_C d^4 x = \int_C\, dt \,
\int_{-\infty}^{\infty} \, d^3{\bf x} $ where $C$ is {\em any} contour
in the imaginary time plane starting at $t_i$ and ending at $t_i -
i \beta$ \cite{LvW,lB}:  here we do not restrict the calculation to
any particular thermal field theory formalism.
In this path integral representation the
$\psivec$'s are now fields (and no longer operators) which are
anti-periodic;
\begin{equation}
\psivec(t,{\bf x}) = - \psivec(t-i\beta,{\bf x}).
\label{anti}
\end{equation}
In the usual way, one can rewrite the quartic term in (\ref{LBCS})
through the identity\tnote{Note that there is no need to assume
that the bosonic $\Delta(t,\vecx)$ fields are periodic even though
the $\psi_{\downarrow} \psi_{\uparrow}$ operator is, not least
because we may couple external, non-periodic sources to the pair
field to set up the initial conditions.}
\begin{equation}
\exp \left[ i g \int_C d^4 x
\psi^{*}_{\uparrow} \psi^{*}_{\downarrow} \psi_{\downarrow} \psi_{\uparrow}
\right]
=
\int \calD \Delta^* \calD \Delta \exp \left[ i \int_C d^4 x \left(
\Delta^* \psi_{\downarrow} \psi_{\uparrow} + \Delta
\psi^{*}_{\uparrow} \psi^{*}_{\downarrow} - \frac{1}{g} |
\Delta|^2 \right) \right]
\label{introD}
\end{equation}
which on substitution into (\ref{partZ}) and
performing the Gaussian integral over the Grassman fields
gives
\begin{equation}
Z =  \int \calD \Delta^{*} \calD \Delta
\exp \left[ i {\Seffbar}[\Delta^*, \Delta]   \right].
\end{equation}
Here
\begin{equation}
{\Seffbar}[\Delta^*, \Delta] = {\Seff}[\Delta^*, \Delta] -
\frac{1}{g} \int_C d^4 x  \;  \Delta^{*}  \Delta
\label{Sbarfund}
\end{equation}
where
the effective action $\Seff$ is given by
\begin{eqnarray}
\Seff[\Delta^*, \Delta]
&=& -i {\rm Tr} \ln \matK^{-1}
 =  -i {\rm tr} \int_C d^4 x \ln
\matK^{-1}(x,x)
\label{Sefffund}
\end{eqnarray}
and the matrix $\matK^{-1}$ by
\begin{eqnarray}
\matK^{-1}(x,y) &=& \left(
\begin{array}{cc}  i \paa_t +
\frac{\nabla^2}{2m} + \mu & \Delta(t, \vecx) \\
\Delta^{*}(t,\vecx) &   i \paa_t -
\frac{\nabla^2}{2m} - \mu
\end{array} \right) \delta^{4}(x-y)  .
\label{Kdef}
\end{eqnarray}
Finally, in equation (\ref{Sefffund}), `tr' means the
trace over matrix components.

The purpose of the remainder of this paper is to calculate $\Seff$.
Such a calculation must clearly be done within an approximation scheme,
and this consists first of a weak field expansion (section
\ref{secwf}) and then a derivative expansion (sections \ref{secde} and
\ref{seccalc}).

\section{The Weak Field Expansion}\label{secwf}

The first component of the approximation scheme is a weak field
expansion.
We work in the symmetry broken phase, so that $0 \leq T < T_{C}$ where
$T_{C}$ is the critical temperature, and then expand
$\Delta(t,\vecx)$ in the effective action $\Seff[\Delta^*, \Delta]$
by writing
\begin{equation}
\Delta(t,\vecx) = \left( | \Delta_0 | + \phi(t,\vecx) \right)
e^{i\theta_0}.
\end{equation}
Here $\Delta_0 = | \Delta_0 | e^{i \theta_0}$ is the $x$ independent
solution of the gap equation
\begin{equation}
\left. \frac{\delta {\Seffbar}}{\delta \Delta^*} \right|_{\Delta=\Delta_0}
=
\left. \frac{\delta {\Seff}}{\delta \Delta^*} \right|_{\Delta=\Delta_0}
- \; \; \frac{1}{g} \; \Delta_0
= 0.
\label{gapdef}
\end{equation}
One can absorb the phase by making a symmetry transformation so
without loss of generality we will take
$\Delta_0 $ real, $\theta_0=0$.

In order to expand about $\Delta_0$, first
introduce the matrices $\matphi(x)$ and $\matG^{-1}(x,y)$
defined by
\begin{eqnarray}
\matphi(x) & = &  \left(
\begin{array}{cc}  0 & \phi(t,\vecx) \\
\phi^{*}(t,\vecx) &  0
\end{array} \right)
\nn
\\
\matG^{-1}(x,y) & = &  \left(
\begin{array}{cc}  i \paa_t +
\frac{\nabla^2}{2m} + \mu & \Delta_0 \\
\Delta_0 &   i \paa_t -
\frac{\nabla^2}{2m} - \mu
\end{array} \right) \delta^{4}(x-y) ,
\label{Ginvdef}
\end{eqnarray}
and also the inverse matrix $\matG(x,y)$ which satisfies
\begin{equation}
\int d^4 y \; \matG^{-1}(x,y) \; \matG(y,z) = \delta^{4}(x-z)
\matone .
\label{propG}
\end{equation}
Now, by definition of $\matK^{-1}$ in (\ref{Kdef}), it follows that
\begin{eqnarray}
\matK^{-1}(x,y) & = & \matG^{-1}(x,y) +
\matphi(x) \delta^{4}(x-y)
\nn
\\
& = & \int d^4 z \; \matG^{-1}(x,z) \left[ \delta^{4}(x-z) \matone +
\matG(z,y)
\matphi(x) \right].
\label{me}
\end{eqnarray}
Hence substitution of (\ref{me}) into (\ref{Sefffund}) followed by an
expansion of the logarithm
in powers of $\phi(x)$ gives
\begin{equation}
\Seff [\Delta^*,\Delta]  =
\Seff^{(0)}[\phi^*,\phi] +
\Seff^{(1)}[\phi^*,\phi] +
\Sefft[\phi^*,\phi] +
 {\cal{O}}(\phi^3)
 \label{star}
\end{equation}
where
\begin{eqnarray}
\Seff^{(0)}[\phi^*,\phi]
&=&
-i {\rm tr} \int_{C} d^4 x \ln \matG^{-1}(x,x)
\nn
\\
\Seff^{(1)}[\phi^*,\phi]
&=&
-i {\rm tr} \int_{C} d^4 x \; \matG(x,x) \matphi(x)
\label{Seff1}
\\
\Sefft [\phi^*,\phi]
&=&
\frac{i}{2}
{\rm tr} \int_{C} d^4 x \int_{C} d^4 y \; \matG(x,y) \matphi(y)
\matG(y,x) \matphi(x) .
\label{Seff2}
\end{eqnarray}
Notice that whilst the first two terms in the expansion of $\Seff$ are
local, the third one $\Sefft$ is non-local, as indeed are the higher
order terms in $\phi$.  The purpose of the derivative expansion, which
is the
second ingredient in the approximation scheme used to calculate
$\Seff$, is to remove this non-locality.  Here we focus on
$\Sefft$, and the derivative expansion is done by Taylor expanding
$\matphi(y)$ in (\ref{Seff2}) as
$$
\matphi(y)= \matphi(x) + (y-x) \frac{\partial \matphi(x)}{\partial x} + \ldots
.
$$
It is
this derivative expansion which brings us into contact with the
technical problem of ill-defined zero energy and momentum limits
of thermal Green functions, and these in turn cause the
inconsistencies usually found in such expansions at finite
temperature \cite{AT,St,vW,AL}.

% ********************************************************************
\section{The derivative expansion}\label{secde}

Before carrying out the above mentioned derivative expansion on
$\Sefft$, we try to summarise the different approaches to this
expansion that have
previously been used in the literature \cite{Popov,AdriaanLH,AT,St,AL}.

First consider the procedure carried out by authors who have worked in
the imaginary time formalism of thermal field theory
\cite{Popov,AdriaanLH,AT,AL}.  Typically their calculations proceed through
the following stages.  Normally the first step is to Fourier transform equation
(\ref{Sefffund}) into momentum space to give
\begin{equation}
\Sefft = -\frac{i}{2} \, {\rm tr} \sum_{k_0}
\int \frac{d^3 \veck}{(2 \pi)^3}
\sum_{p_0}
\int \frac{d^3 \vecp}{(2 \pi)^3} \;
\matphi (p) \matG(k-p) \;
\matphi(-p) \; \matG(k).
\label{mmtSeff}
\end{equation}
A simple but important point to note here is that the energies,
$k_0,p_0$, have been taken to be imaginary and discrete as is typical for
systems in equilibrium.
The second step is to expand expression
(\ref{mmtSeff}) about $p =0$ in powers of $p_0$ and $\vecp$,
and then evaluate the sums and integrals over $\veck$.  Finally, the
resulting expressions are converted back into real times (that is
some analytic continuation is carried out) to give an expression for
$\Sefft$ in terms of derivatives of $\phi(x)$.
Notice that in the first step of these
calculations---that is, in writing down equation
(\ref{mmtSeff}) with the discrete momenta---both space and time
translation invariance have been assumed, and furthermore the fields and
propagators involved in the rest of the calculation are those of {\em
equilibrium systems}.  It follows that in the second step,
discrete momenta (or equivalently
equilibrium fields) have been used within the derivative expansion
itself---below we will comment further on this point.

The aim of this section is to discuss whether or not the operations
outlined in the above paragraph are in fact consistent.

To that effect consider now expression (\ref{Seff2}) in $(t,\vecx)$
space (rather than the Fourier transformed expression (\ref{mmtSeff})).
In particular, focus on the time component of (\ref{Seff2})
since it is this which changes between the $T=0$ and $T>0$
calculation, and also let $x=(t,\vecx)$ and $y=(t',\vecy)$.  The
derivative expansion of $\Sefft$ is obtained by writing
\begin{equation}
\matphi(t')= \matphi(t) + (t'-t) \frac{\partial \matphi(t)}{\partial
t} + \ldots
\label{order1}
\end{equation}
Substitute (\ref{order1}) into (\ref{Seff2}) and consider the
contribution to $\Sefft$ which arises say from the term containing the
time derivative $\partial_t \matphi(t)$.  Clearly this term is {\em
not} periodic in $t'$ in the imaginary time direction since
$t'-t-i \beta \neq t'-t$.  Indeed this conclusion holds for {\em
every} individual term in the derivative expansion of
(\ref{Sefffund}), apart from the $t'$-independent contribution.

Thus we arrive at the conclusion that {\em each term in the derivative
expansion of (\ref{Seff2}) is not periodic in imaginary
time}.  (One should perhaps
not be surprised at this---by the very essence of
the derivative expansion one is dealing with slowly time varying
and therefore non-equilibrium systems.)
Hence it follows that at each order in the derivative expansion,
taking a Fourier transform would {\em not} give rise to discrete
energies, and hence it is {\em
not} correct to use discrete energies (which correspond to periodic
fields) if one were to work with
Euclidean energies.  However, we have stressed in the first paragraph
of this section that authors who previously calculated derivative
expansions in the ITF formalism {\em did} assume such discrete energies.
We therefore believe
that these previous calculations are inconsistent.

Before suggesting a method to overcome this inconsistency, note that
the above discussion has emphasised the imaginary time approach to thermal field
theory whilst some authors have used a real time approach \cite{St}.
Recall though that these alternative real-time methods always give
the same answer as the imaginary time calculations for any
physical quantity \cite{BM,FW,TSEnpt}.  In particular, this is true
when one compares
results for thermal effective actions, provided note is taken of
various technical issues \cite{We,EP,lBM}.
We now rephrase the above arguments in a way
suitable for real-time methods.  This is done by discussing the
periodicity of the fields---a boundary condition which is common to
all path-integral approaches to thermal field theory.
The generalisation of
expression (\ref{order1}) to infinite order is
\begin{equation}
\label{phittrans}
\matphi(t')= \left. e^{i E'' (t'-t) }\matphi(t'')  \right|_{t'' =
t}
\end{equation}
where
$$
E''=-i \frac{\partial}{\partial t''}.
$$
If the left hand side, $\matphi(t')$, is periodic in imaginary time,
then for equality the right hand side must also be periodic.  This
is assumed in previous calculations \cite{St} and is clearly
only true if $e^{\beta
E''}=1$, a condition which was imposed in those calculations.
However, as was noted above, each individual term in the derivative
expansion (in other words the expansion of the exponential in
\tref{phittrans}), 
is not periodic.  Furthermore, since one always truncates the
series and works to a finite order in derivatives, it follows that the
approximation to the
field $\matphi(t)$ used in the expansion
{\em cannot} be periodic. Thus, we believe that
setting $e^{\beta E''}=1$ is {\em not} consistent with the
derivative expansion. Again, we stress that since this was done in
previous calculations, we believe these to be inconsistent.

The approach to the derivative expansion presented in this paper
therefore has one crucial difference with previous calculations.  As
shown above, each term in the derivative expansion is not
periodic, and therefore our calculation does not assume periodicity.
This is equivalent to saying that we allow $e^{\beta E''}
\neq 1$, or equivalently that we do not work with discrete
momenta for the field $\matphi(t)$---as is typical of non-equilibrium
systems.
Given the lack of discrete momenta, it is difficult to
work in $(p_0,\bfp)$ space and instead we work in $(t,\bfp)$
space.
This is in fact a
great simplification as it turns out that all the time
integrations can be done exactly for any {\em arbitrary} thermal field
theory formalism, and thus it avoids all the above
mentioned issues.   Furthermore, there are no problems with
analytic continuation, which is another point we feel has been
passed over rather rather hastily in the existing
calculations.
Finally, note that consistently with these discussions, we obtain
the same results as in the existing literature \cite{AT,AL} if we set
$e^{\beta E''}=1$.\footnote{Regarding the discussions of
(anti-)periodicity made above, it should be noted that $G(t,\veck)$
is the propagator for the Fermi fields, which by definition in equation
(\ref{anti}) {\em are} anti-periodic in imaginary time.  Hence $G(t,\veck)$
{\em is} anti-periodic; $G(t,\veck)=-G(t-i\beta,\veck)$.
Regarding $\matphi(t,\veck)$, or
equivalently the field $\Delta(t,\veck)$, we have shown that for the
derivative expansion to be consistent at any finite order,
$\Delta(t,\veck)$ {\em cannot} be periodic.
That this is the case could be achieved by adding a small time dependent
perturbation to $\Delta$, and we are then watching the system respond
to this.  In
addition, it should be noted that in equation (\ref{introD}) which
introduces the $\Delta(t,\veck)$ field, there is nothing to constrain
$\Delta$ to be periodic in the first place.}

Our convention for Fourier
transforms is
$$
f(t,\bfx)  = \intdtk
f(t,\veck) e^{-i \veck.\bfx}.
$$

% ***********************************************************
\section{The calculation}\label{seccalc}

For the reasons discussed in the previous section, namely that each
term in the derivative expansion leads to non-periodic functions, we
work throughout in $(t,\veck)$ space.  The first part of the
calculation, section \ref{subsecprop}, consists of determining the
propagator $\matG(t,t',\veck)$ which appears in $\Sefft$.
The gap equation is then
calculated in section \ref{subsecgap} whilst the derivative expansion
of $\Sefft$ is set up in section \ref{subsecderiv}.  There we will see
that by working with the time variables, the integrations can indeed be done
explicitly for an arbitrary thermal
field theory formalism.  At the same time this avoids the issues of
analytic continuation which complicate some approaches
\cite{BM,TSEnpt,TSEllwi}.

\subsection{The fermion propagator}
\label{subsecprop}

As was discussed in section \ref{secde}, the fermion propagator
$\matG(t,t',\veck)$ is in fact an equilibrium propagator since the Fermi
fields are in equilibrium by equation (\ref{anti}).  Its calculation
therefore follows standard methods which we outline
here; the details are given in appendix \ref{appdp}.

First use (\ref{Ginvdef}) to define the matrix $\matG^{-1}(x)$ through
$$
\matG^{-1}(x) \delta^4(x-y) := \matG^{-1}(x,y)
$$
so that in $(t,\bfk)$ space
\begin{equation}
\matG^{-1}(t,\bfk) =
\left(
\begin{array}{cc}  i \paa_t -
\frac{\bfk^2}{2m} + \mu & \Delta_0 \\
\Delta_0^{} &   i \paa_t +
\frac{\bfk^2}{2m} - \mu
\end{array} \right)
=
\left(
\begin{array}{cc}  i \paa_t -
\epsk & \Delta_0 \\
\Delta_0^{} &   i \paa_t +
\epsk
\end{array} \right)
\label{GG}
\end{equation}
where $\epsk$ is the energy measured from the Fermi surface;
$$
\epsk = \frac{\veck^2}{2m} - \mu.
$$
As usual in the BCS limit in which we are working here, the Fermi
wavenumber is defined through the chemical potential $\mu$ by $k_F^2 =
2 m \mu$.

Now, from equation (\ref{propG}), $\matG(t,t',\veck)$ satisfies
$$
\matG^{-1}(t,\bfk) \matG(t,t',\bfk) =  \delta(t-t') \matone.
$$
In the usual way one
can solve this equation for $\matG(t,t',\veck)$ by diagonalising
the operator  $\matG^{-1}(t,\bfk)$ which given in (\ref{GG}) (see
appendix \ref{appdp}).  The result is
\begin{equation}
\matG(t,t',\bfk) =  \left(
\begin{array}{cc}
u^2_+ F_-(t,t',\bfk) + u^2_- F_+(t,t',\bfk) & u_+
u_- (F_+(t,t',\bfk)
- F_-(t,t',\bfk) ) \\
u_+ u_- (F_+(t,t',\bfk)- F_-(t,t',\bfk))  &
u^2_+ F_+(t,t',\bfk) +
u^2_- F_-(t,t',\bfk)
\end{array} \right).
\label{AnswerG}
\end{equation}
Here $u_\pm$ are functions of $\veck$ (this dependence has not been
written in (\ref{AnswerG}) for clarity), and are given by
$$
u_\pm(\veck) = +\left( \half \left[ 1 \pm
\frac{\epsk}{\Ek} \right] \right)^\half
$$
where $\Ek$
is the dispersion relation
$$
\Ek = + \left[ |\Delta_0|^2 + \epsilon^2_\veck \right]^\half.
$$
It is important to recognise that $u_+(\veck)$ and $u_-(\veck)$
are just the usual functions often called $u(\veck)$ and
$v(\veck)$ respectively which, in the operator approach, are
the coefficients of the
Bogoliubov transformation which diagonalises the Lagrangian.
We have chosen this different notation as later (in section
\ref{seccalc}) it will enable us
to write our results in a significantly more compact form.

As a result of the diagonalisation procedure (appendix \ref{appdp}),
the functions $F_\pm(t,t',\bfk)$ which appear in (\ref{AnswerG}) are
seen to satisfy the Feynman equation
$$
(i \partial_t \pm \Ek) F_\pm(t,t',\bfk) = \delta(t-t') \matone.
$$
Hence their solutions are known, and for our purposes they are
best given in a Mills
representation
\cite{Mills,lBM};
\begin{eqnarray}
F_\pm (t,t',\veck) &=& -i \int
\frac{dk_0}{2 \pi} e^{-i k_0 (t-t') } \left( \theta_C(t,t') -
N(k_0) \right) \rho_\pm(k_0,\veck).
\label{Fdef}
\end{eqnarray}
Here $N(k_0)$ is the Fermi-Dirac distribution:
$$
N(k_0) = \frac{1}{e^{\beta k_0} + 1},
$$
the $\theta_{C}$ is the usual $\theta$ function for the contour $C$
\cite{LvW,lB},
and the spectral functions $\rho$ are given by
\begin{equation}
\rho_\pm(k_0,\veck) = 2\pi \delta(k_0 \pm \Ek).
\label{rho}
\end{equation}
Finally we note that $F_+(t,t',\veck) = F_-(t',t,\veck)$.

\subsection{The Linear terms and the Gap Equation}
\label{subsecgap}

Before turning to the calculation of $\Sefft$ in (\ref{Seff2}), we
consider briefly the gap equation (\ref{gapdef}) or in other words the
terms linear in $\phi$ in $\bar{S}_{\rm eff}$.  As usual the gap
equation is a necessary tool in the evaluation of $\Sefft$ as it is
used to remove UV divergences (see below).

Equation (\ref{Seff1}) gives $\Seff^{(1)}$ to be
$$
\Seff^{(1)}[\Delta^*,\Delta]
=
- i \int_{C} d^4 x
\left(  {G}_{12}(0) {\phi}^*(x) +
{G}_{21}(0) {\phi}(x) \right)
$$
where from (\ref{AnswerG})
\begin{eqnarray}
{G}_{12}(0) :=
{G}_{12}(x,x) &=&
i \intdtk
 \frac{\Delta_0 }{2 \Ek}
(F_+(t,t,\veck) - F_-(t,t,\veck))
\nn
\\
&=&
- i \intdtk
 \frac{\Delta_0 }{2 \Ek}
(1-2N(\Ek) ).
\nn
\end{eqnarray}
Thus on substitution into (\ref{gapdef}) one obtains
\begin{equation}
\intdtk
 \frac{1 }{2 \Ek}
(1-2N(\Ek) )
= - \; \frac{1}{g}
\label{gap111}
\end{equation}
which agrees, for example, with Stoof \cite{St}.  Notice that the
first term on the left hand side of (\ref{gap111}) is linearly
divergent since $\Ek \propto \veck^2$ for large $\veck$.

% *****************************************************************

\subsection{Quadratic Terms}
\label{subsecderiv}

Now consider the important quadratic term $\Sefft$ of the
effective action (\ref{Sefffund}).  Substitute  the propagator
(\ref{AnswerG}) into (\ref{Seff2}) Fourier transformed into
$(t,\veck)$ space.  On taking the trace in (\ref{Seff2}) one obtains
four terms.
Two of these are proportional to
$\phi^*(t,\bfp) \phi(t',-\bfp)$ and $ \phi(t,\bfp) \phi^*(t',-\bfp)$
and, as is common in the literature, we call these the diagonal terms.
The non-diagonal terms are proportional to $\phi^*(t,\bfp) \phi^*(t',-\bfp)$
and $\phi(t,\bfp) \phi(t',-\bfp)$ and in fact the coefficients of
these two non-diagonal terms will be seen to be identical.  Thus we write
\begin{equation}
\Sefft =
 \Sefft[{\rm Da}] + \Sefft[{\rm Db}] + \Sefft[{\rm ND}]
\label{decomp}
\end{equation}
where the D and ND stand for diagonal and non-diagonal as explained above.
The diagonal terms are given by
\begin{eqnarray}
\Sefft[{\rm Da}]
&=&
\frac{i}{2} \int_{C} \, dt \int_{C} \, dt' \intdtp \intdtk
\phi(t,\bfp) \phi^*(t',-\bfp) \times
\nn
\\
&& \times
\left[ u_+^2(\bfkmp)
F_+(t,t',\bfkmp) +   u_-^2(\bfkmp) F_-(t,t',\bfkmp)
\right]
\times
\nn
\\
&& \times
 \left[ u_+^2(\bfkpp) F_-(t',t,\bfkpp) +
 u_-^2(\bfkpp) F_+(t',t,\bfkpp) \right]
\label{Da}
\end{eqnarray}
and
\begin{eqnarray}
\Sefft[{\rm Db}]
&=&
\frac{i}{2} \int_{C} \, dt \int_{C} \, dt' \intdtp \intdtk
 \phi^*(t,\bfp) \phi(t',-\bfp) \times
\nn
\\
&  & \times
\left[ u_+^2(\bfkpp) F_+(t',t,\bfkpp)
+   u_-^2(\bfkpp) F_-(t',t,\bfkpp) \right]
\nn
\\
&& \times
\left[ u_+^2(\bfkmp)
F_-(t,t',\bfkmp) +   u_-^2(\bfkmp) F_+(t,t',\bfkmp)
\right].
\label{Db}
\end{eqnarray}
Observe that interchange of $t$ and $t'$ in (\ref{Da}) followed by
the relabelling $\bfp \rightarrow -\bfp$ yields $\Sefft[\rm Db] =
\Sefft[\rm Da]$.
However, consider now any given term in the derivative expansion
of $\Sefft[{\rm Da}]$, which is obtained by writing
$\phi^*(t',-\vecp)=\phi^*(t,-\vecp) + (t'-t) \partial_t
\phi^*(t,-\vecp) + \ldots$ in (\ref{Da}).  It is straightforward
to see that the changes $t' \leftrightarrow t$ and $\vecp
\rightarrow -\vecp$ for that term do {\em not} yield the
corresponding term of $\Sefft[{\rm Db}]$ (the only exception is
the $t'$ independent contribution).  That is, whilst this change
of variables will yield $\Sefft[\rm Db] = \Sefft[\rm Da]$ to
infinite order in the derivative expansion, this is not true to
any finite order. Since the derivative expansion is always
truncated, it is therefore necessary to work with $\Sefft[\rm Db]$
and $\Sefft[\rm Da]$ separately. Below we will indeed verify that
the results obtained in this way differ from those of a derivative
expansion on $2\Sefft[\rm Db]$ (see equation \tref{gammaone}).

The non-diagonal term in (\ref{decomp}) is given by
\begin{eqnarray}
\Sefft[{\rm ND}]
&=&
\frac{i}{2} \int_{C} \, dt \int_{C} \, dt' \intdtp \intdtk
\left\{ \phi^*(t,\bfp) \phi^*(t',-\bfp) + \phi(t,\bfp) \phi(t',-\bfp) \right\}
\nn
\\
&& \; \; \; \; \; \times \;
u_+(\bfkmp)u_-(\bfkmp)u_+(\bfkpp)u_-(\bfkpp)
\nn
\\
& & \; \; \; \; \; \times \;  \left[ F_+(t',t,\bfkpp) -   F_-(t',t,\bfkpp) \right]
\nn
\\
&& \; \; \; \; \; \times \;
\left[ F_+(t,t',\bfkmp) - F_-(t,t',\bfkmp)  \right].
\label{ND}
\end{eqnarray}

\subsubsection{Derivative expansion of $\Seff [{\rm Da}]$}

As a specific example we now carry out a derivative expansion on
$\Seff [{\rm Da}]$ given in \tref{Da}.   The computation of
 $\Seff [{\rm Db}]$ and $\Seff [{\rm ND}]$  will
follow very similar steps.

First Taylor expand $\phi(t',-\bfp)$ about $t'=t$;
\begin{eqnarray}
\phi(t',-\bfp) & = & \phi(t,-\bfp) + (t'-t) \left. \frac{\paa
\phi(t,-\bfp)}{\paa t}  \right|_t + \frac{1}{2!} (t'-t)^2 \left.
\frac{\paa^2(t,-\bfp) \phi}{\paa t^2}
 \right|_t
 + \ldots
\label{exp1}
\\
& = & \left. e^{i E'' (t'-t) }\phi(t'',-\bfp)  \right|_{t'' = t}
\label{exp2}
\end{eqnarray}
where $E'' = -i \partial_{t''}$.  It is now possible to proceed in
two different but equivalent ways. The first is to consider a
given term in (\ref{exp1}) (for example one proportional to
$\partial_t^n \phi$), substitute it into (\ref{Da}) and then
calculate its contribution to the effective action.  Equivalently
one can substitute (\ref{exp2}) directly into (\ref{Da}), and then
obtain the contribution from $\partial_t^n \phi$ by simply
calculating the term proportional to $E''^n$ in the expansion of
$\Sefft[{\rm Da}]$ in powers of $E''$.  We follow this second more
powerful method here.

Therefore, substitute (\ref{exp2}) into \tref{Da} to obtain
$$
\Sefft [Da] = \left. \int_C \; dt \intdtp \phi^*(t,\vecp)
\calB_{Da}(t,t_i,t'',\vecp) \phi(t'',-\vecp)
 \right|_{t'' = t}
$$
where $\calB_{Da}$ is given by
\begin{eqnarray}
\lefteqn{ \calB_{Da}(t,t_i,t'',\vecp) :=
\frac{i}{2} \intdtk  \int_{C} \, dt'
\; \; e^{i E'' (t'-t)} \; \; \times}
\nn
\\
& \times & \left[  u_+^2(\bfkmp)  u_-^2(\bfkpp)F_+(t,t',\bfkmp)
F_+(t',t,\bfkpp)
  \right.
\nn
\\
& & + \;     u_-^2(\bfkmp)  u_+^2(\bfkpp)  F_-(t,t',\bfkmp) F_-(t',t,\bfkpp)
\nn
\\
&&
+   \; u_+^2(\bfkmp)  u_+^2(\bfkpp)  F_+(t,t',\bfkmp)F_-(t',t,\bfkpp)
\nn
\\
&&
 \left. + \;  u_-^2(\bfkmp)  u_-^2(\bfkpp)  F_-(t,t',\bfkmp)F_+(t',t,\bfkpp)
\right]
\nn
\\
&=&
\frac{i}{2} \intdtk \sum_{s_0,s_1 = \pm }
u^2_{-s_0}(\bfkmp) u^2_{-s_1}(\bfkpp)
\Zbar_{s_0 s_1}(t,t_i,E'',\veck,\vecp)
\label{BDa}
\end{eqnarray}
and
\begin{equation}
\Zbar_{s_0 s_1}(t,t_i,E'',\veck,\vecp) := \int_C \; dt' \;
F_{-s_0}(t,t',\bfkmp) F_{s_1}(t',t,\bfkpp) e^{i E'' (t'-t)}.
\label{Zbar}
\end{equation}
In traditional work \cite{AT,AL}, one finds that the retarded
thermal Green function \cite{BM,FW} appears instead of
$\calB_{Da}$.  These are the same only in the limit $e^{\beta
E''}=1$ in accordance to the discussion in section \ref{secde}.
 Below we will see
that the so called Landau damping contributions come from the
terms in (\ref{BDa}) with $s_0=-s_1$, and furthermore that they do
not give rise to any problems with the derivative expansion.

The only dependence on the variable $t'$ in \tref{Da} is now in
$\Zbar$, and we show that this may be evaluated for {\em any}
contour $C$.  Thus our method includes both imaginary- and
real-time approaches to thermal field theory. Substitute
expressions \tref{Fdef} for $F_\pm$ into \tref{Zbar} to obtain
\begin{eqnarray}
\Zbar_{s_0 s_1}
& = &
\int_C \; dt'
\int \frac{dk_0}{2 \pi}
\int \frac{dk_1}{2 \pi} \;
\rho_{ - s_0}(k_0,\bfkmp)
\rho_{   s_1}(k_1,\bfkpp)
\times
\nn
\\
& \times &
\left[ \theta_C(t',t) - N(k_1) \right]
\left[ \theta_C(t,t') - N(k_0) \right]
e^{i E'' (t'-t)} e^{i k_0 (t'-t)} e^{-i k_1 (t'-t)}
\nn
\\
& =: &
\int \frac{dk_0}{2 \pi} \; \int \frac{dk_1}{2 \pi}
\rho_{ - s_0}(k_0,\bfkmp)
\rho_{   s_1}(k_1,\bfkpp)
J_{C}(k_0,k_1,t_i-t,E'').
\label{Zdef}
\end{eqnarray}
Observe that all the time dependence is in the function
$J_{C}(k_0,k_1,t_i-t,E'')$ which, on letting $\Abar = E''+ k_0 - k_1$, is given by
\begin{eqnarray}
J_{C}(k_0,k_1,t_i-t,E'') &=& \int_C \; dt'  \;
\left[ \theta_C(t',t) - N(k_1) \right]
\left[ \theta_C(t,t') - N(k_0) \right]  e^{i\Abar (t'-t)}
\label{JCdef}
\nn
\\
& = &
\int_C \; dt' N(k_0)N(k_1) e^{i \Abar
(t'-t)} -
N(k_1) \int_C \; dt' \theta_C(t,t')
e^{i \Abar (t'-t)} 
\nn
\\
& - & N(k_0) \int_C \; dt' \theta_C(t',t)
e^{i \Abar (t'-t)}
+ \int_C \; dt' \theta_C(t',t)\theta_C(t,t')e^{i \Abar (t'-t)}.
\nn
\end{eqnarray}
Consider now an {\em arbitrary} contour
$C$ starting at $t_i$ and ending at $t_i - i \beta$.  The
integrals in (\ref{JCdef})
may be calculated explicitly for that contour $C$:  using
the definition of $ \theta_C(t',t)$, the final term contributing to
$J_{C}(k_0,k_1,t_i-t,E'')$ vanishes; the only contribution to the second
term is for $t_i < t' < t$; and the contribution to the third term is
for $t < t' < t_i - i \beta$.  Thus the second term, for example, is
given by
\begin{eqnarray}
N(k_1) \int_C \; dt' \theta_C(t,t')  e^{i \Abar
(t'-t)} & = &
N(k_1)e^{-i \Abar t}  \int_{t_i}^t \; dt' \; e^{i \Abar t'}
\nn
\\
& = &
\frac{1}{i\Abar} N(k_1)e^{-i \Abar t} \left[ e^{i \Abar t} - e^{i \Abar t_i}
\right].
\nn
\end{eqnarray}
After similar manipulations for the other terms, we obtain
\begin{equation}
J_{C}(k_0,k_1,\delta,E'') = - \frac{N(k_0)N(k_1)}{i \Abar} \left[
e^{\beta k_1} \left( e^{\beta \Abar} - 1 \right) + e^{\beta k_0} \left(
1 - e^{i \Abar \delta} \right) \left( 1 - e^{\beta E''}  \right) \right]
\label{JC1}
\end{equation}
where $\delta = t_i - t$.  This step is perhaps the most important
one of our method---by working with the time variable $t$ itself
rather than Fourier transforming, we have been able to do the
integrals explicitly and without using the explicit form for the
spectral functions $\rho$.

Substituting (\ref{JC1}) back into (\ref{Zdef}) and using the
definitions of $\rho$ in (\ref{rho}) gives
\begin{eqnarray}
\lefteqn{\Zbar_{s_0s_1} (t,t_i,E'',\veck,\vecp)
 =}
\nn
\\
&&\frac{1}{(\exp \{ \beta s_0 E_\bfkmp\} +1)}
\frac{1}{(\exp\{ \beta s_1 E_\bfkpp\} +1)}
\frac{1}{i A}
\nn
\\
&&
\times
\left\{ \left( 1 - e^{\beta A} \right)
 + \exp \{ \beta (s_0 E_\bfkmp + s_1 E_\bfkpp \}
 \left( e^{i \delta A} -1 \right)
 \left( 1 - e^{\beta E''} \right) \right\}
\label{Zbar1}
\end{eqnarray}
where we have used the shorthand notation
\begin{eqnarray}
A &=& E''+ s_0 E_\bfkmp + s_1 E_\bfkpp.
\label{shorthand}
\end{eqnarray}

Before going any further we make the following comments.
First, observe that if we set $e^{\beta E''}=1$ then our
expression (\ref{Zbar1}) reduces to the corresponding equation
(number (10)) of \cite{AL}.  This is consistent with the
discussion of section \ref{secde}.

Secondly, note that the time independent contributions to the
effective action have $E''=0$ and hence $e^{\beta E''}=1$.  Thus
all our time independent results will be identical to those of
\cite{St,AL}.

Thirdly, equation (\ref{Zbar1}) has an explicit dependence on
the initial time $t_i$.  Time translation invariance has therefore been
lost as expected for such a time dependent non-equilibrium system.
Furthermore, an important point to observe is that even at $t=t_i$
($\delta=0$), our expression \tref{Zbar1} differs similar terms in
\cite{AT,AL} by a factor of $e^{ \beta E'' }$.  This point
will be discussed further below.

Finally, consider the terms with $s_0=-s_1$ in
$\calB_{Da}$.  From equations (\ref{BDa}) and
(\ref{Zbar}) and the expression for $\Zbar$ in (\ref{Zbar1}), these 
terms correspond to Landau damping terms in the original
formalism (when $e^{\beta E''} = 1$).  The reason is that in this
case, the denominators 
contain $E'' \pm ( \Ekmp - \Ekpp)$, which vanishes when $\bfp=0$
and $E''=0$ thus leading to a potential divergence.  It is this
divergence in the bubble diagram which is sometimes ignored
\cite{St}, and was the cause of the badly defined derivative
expansion in \cite{AL}.  However, observe that in our case we have
$e^{\beta E''} \neq 1$, so that {\em the} $\bfp=0$ and $E''=0$ {\em 
limits are
actually well defined}:  the zero in the denominator $\calB_{Da}$ when
$s_0=-s_1$ is exactly cancelled by a zero in the
numerator of this expression. Thus here we are dealing with
Landau damping terms that have a well defined derivative
expansion.  Similar comments will apply for the other diagonal term
as well as the non-diagonal terms which we now 
calculate.\footnote{Interestingly, the cuts at $E'' = \pm 2
\Delta_0$ familiar from Green functions commonly encountered in
field theory are also not present in the $\calB$'s when $s_0=s_1$.
However, one must be very precise when making the link between
physics and the analytic properties of particular Green functions.
For instance it is the theory of linear response which links cuts
at $E'' = \pm 2 \Delta_0$ in retarded Green functions to physical
decay processes at those energy scales \cite{lB,FW}.  Here we are
{\em not} using linear response theory but rather looking at
systems which are slowly varying in time.  It is therefore not
clear that $\calB$ should have the same analytic structure as the
retarded Green functions of linear response \cite{lB,FW} as they
found to be relevant to different physical problems.}

\subsubsection{Second diagonal term $\Sefft[{\rm Db}]$}

Calculation of $\Sefft[{\rm Db}]$ follows a very similar route to
that above.  Indeed, the function $\Zbar$  introduced above also
appears in $\Sefft[{\rm Db}]$ as can be seen by comparing the
temporal dependence of $\Sefft[{\rm Da}]$ in (\ref{Da}) with that
of $\Sefft[{\rm Db}]$ in (\ref{Db}).  This leads to a switch from
$u_-$ to $u_+$ functions and one obtains
$$
\Sefft[{\rm Db}]  =   \left. \int_C dt \intdtp {\phi}^*(t,\bfp) {{\cal
B}}_{Db}(t,t_i,t'',\vecp) {\phi}(t'',-\bfp)
 \right|_{t''=t}
$$
where the bubble diagram ${{\cal B}}_{Db}(t,t_i,t'',\vecp)$
contains contributions from all the four terms in (\ref{Db}):
\begin{equation}
\label{BDb} {{\cal B}}_{Db} = \frac{i}{2} \intdtk
\sum_{s_0,s_1=\pm} u^2_{s_0}(\bfkmp) u^2_{s_1}(\bfkpp) \Zbar_{s_0
s_1}.
\end{equation}

\subsubsection{Non-diagonal term $\Sefft[{\rm ND}]$}

The non-diagonal terms  (\ref{ND}) again follow a similar pattern.
Here, since
$$
u_{+}(\bfkmp) u_{+}(\bfkmp) u_{-}(\bfkpp) u_{-}(\bfkpp)
= \frac{|\delo|^2}{4 \Ekpp \Ekmp}
$$
it follows that
$$
\Sefft[{\rm ND}]  =   \left. \int_C dt \intdtp {\phi}^*(t,\bfp) {{\cal
B}}_{ND}(t,t_i,t'',\vecp) {\phi}(t'',-\bfp)
 \right|_{t''=t}
$$
where the bubble diagram ${{\cal B}}_{ND}(t,t_i,t'',\vecp)$ is
\begin{equation}
{{\cal B}}_{ND} =
\frac{i}{2}
\intdtk
 \frac{|\delo|^2}{4 \Ekpp \Ekmp}
\sum_{s_0,s_1=\pm} (-s_0 s_1) \Zbar_{s_0 s_1}. \label{BND}
\end{equation}

\subsubsection{Total quadratic term and derivative expansion}

To conclude, we have
$$
 \Sefft = \Sefft[{\rm Da}] + \Sefft[{\rm Db}] +
\Sefft[{\rm ND}]
$$
where
$$
\Sefft = \int_{c}dt \int \frac{d^3\bfp}{(2 \pi)^3}
\left.
( \phi^*(t,\bfp) \; \; \; \phi(t,\bfp) )
\left( \begin{array}{cc}
\calB_{Da} & \calB_{ND} \\
\calB_{ND} & \calB_{Db}
\end{array} \right)
\left( \begin{array}{c}
\phi^*(t'',-\bfp) \\ \phi(t'',-\bfp)
\end{array} \right)
\right|_{t''=t_i}
$$
with the $\calB$'s given in \tref{BDa}, \tref{BDb} and
\tref{BND}.

It is now straightforward to expand this effective action
in powers of $\paa_t^n \phi$ and $\nabla^n \phi$ where $n \geq 0$; 
it is done by expanding the $\calB$'s above about $E''=0$ and
$\vecp=0$.\tnote{One can then obtain the equations of motion.}
We now carry out this derivative expansion.

\section{The derivative expansion in the degenerate gas
approximation}\label{secres}

Though it is not difficult to expand the expressions given above in
$E''$ and $\bfp$ to obtain the derivative expansion and hence
effective equations of motion, the resulting expressions are
however very complicated.  To simplify them
we therefore work in the degenerate gas approximation
\cite{Popov} for which $T < T_C$, and furthermore the
gap $\Delta_0$ is much bigger
than the temperature.

Before clarifying further this approximation, note that our
expansion also imposes $\beta |\paa_t \ln(\phi)| \ll
1$ since the factors of
$\exp \{ \beta \Edd \} \equiv \exp \{ -i \beta \paa_{t''} \}$ must be
expanded.  It is fundamental to remember that by their very
definition, derivative expansions are valid only for fields
varying slowly in space and time, and we should not be surprised
that our expressions differ from those used in linear response
theory.  The latter uses retarded Green functions which describe
the response to a delta-function impulse, a very different
physical problem.  Also with $| \partial_t{\phi}/\phi | \ll T$ for
the expansion to be valid, clearly taking the zero temperature $
T\rightarrow 0$ limit makes no sense in our expressions.

To summarise, the results presented below are valid in the regime
$$
  \mu \gg \Delta_0 \gg \beta^{-1} = T \gg | \partial_t{\phi}/\phi
  |,
$$
and the gap $\Delta_0$ will be used as the primary mass scale for the problem.
Thus we
introduce the following dimensionless parameters
$$
y= \frac{\epsilon}{\Delta_0}, \; \; \; z^2 = (1+y^2)=
\frac{E^2}{\Delta_0^2}  .
$$
To exploit the degenerate gas approximation, observe
that the integrals are sharply peaked around the Fermi surface
$k=k_f$. This is not true if there are UV divergent terms but we
shall see that in the effective potential, these term cancel through the gap
equation. The Fermi-Dirac function $N_f$ then appears as
\[
N_f= \left[ \exp \{ \beta \Delta_0 z  \} +1 \right]^{-1} ,
\]
and is even more sharply peaked than other parts of the expression.
Thus we perform the following simplifications:-
\begin{itemize}
\item Isolated factors of $k$ not inside an $\epsk$ or $\Ek$ factor
can be replaced by a constant $k_f$.  Valid since
$k^2=k_f^2(1+y^2/Y^2)$ where $Y:= \mu /\Delta_0 \gg 1$.
\item In terms with {\em at least one} factor of $N_f$,
$\epsk \approx 0, \Ek \approx \Delta_0$, i.e.\ $y \approx 0, z \approx 1$.
Valid since $\Delta_0 \gg T$.
\end{itemize}
The results then have a common factor of $\rho$
\[
\rho := \frac{m k_f}{2 \pi^2} .
\]
For example, the gap equation in this notation is
$$
-\frac{1}{g} = \rho \int dy \; \left( \frac{1}{2z} - N_f \right) .
$$
The expressions will also depend on the Minkowskii time $t$ elapsed
since the initial conditions were set at time $t_i$.

\subsection{Results}\label{results}

The notation in this section is as follows.  From equations
(\ref{Sbarfund}) and (\ref{star}), the effective action is given
by\tnote{Factor of 2 and nature of complex conjugate terms?  Note
in particular that the Higgs/Goldstone split may depend crucially
on this. My $\gamma$ is Dani's $\calB$.  There is a minus sign
from the fermionic trace in the $\Seffbar$ to $\Seff$ relation,
and then there is the $(-i)^2$ factor added to Dani's $F$'s which
is in the $\Seff$ to $\calB$ relation.}
\begin{eqnarray}
\Seffbar &:=& - i {\rm tr} \intdfx \ln \matr{G}^{-1}(x,x) -
\frac{1}{g} \intdfx \Delta_0^2
 \nn
 \\
&+& \intdfx \left(
  \phi^* \gamma^{Da} \phi
+ \phi   \gamma^{Db} \phi^*
+ \phi^* \gamma^{ND} \phi^*
+ \phi   \gamma^{ND} \phi
\right)
\label{ane}
\end{eqnarray}
where $\phi=\phi(t,\vecx)$, $\phi^*=\phi(t,\vecx)$ and $\gamma$'s
are of the form
\begin{equation}
\gamma^X = \sum_{a,b} \gamma^X_{a,b} \left( -i
\frac{\partial}{\partial t} \right)^a \left( i \vecnab \right)^b.
\label{gammaexp}
\end{equation}

The zeroth order effective potential terms, $\gamma^X_{0,0}$ are
easily obtained by first setting $E''=0=\vecp$ in (\ref{Zbar}) and
then using the gap equation to remove the UV divergences.  In the
degenerate gas approximation we have\tnote{$\gamma = -R7d/2$ of the
Maple programmes, the results after DEG has been imposed as that
then catches the $4\pi^2$ factor.
The following results
come from MAPLE programme, {\tt aitchanal5.mws}, which analyses
the results in {\tt R7dresults.map} produced by the latest {\tt
expa21.mws}.}
\begin{eqnarray}
\gamma^{{\rm Da}}_{0,0} & = &  \gamma^{\rm Db}_{0,0} = \gamma^{\rm ND}_{0,0}
\nn
\\
& = & \frac{ \rho }{8} \int dy \left[  \,{\frac {1}{{z}^{3}}}
- 2\Nf - 2(\Delta_0 \beta) N_f (1-N_f) \right] .
\end{eqnarray}
The first order terms are slightly more complicated in part and
are given by
\begin{eqnarray}
\gamma^{\rm Da}_{0,1} & = &  \gamma^{\rm Db}_{0,1} \; \; \; = \;
\; \; \gamma^{\rm ND}_{0,1} \; \; \; = \; \; \; 0, \nn
\\
\gamma^{\rm Da}_{1,0}
& = &
-\frac{ \rho \beta}{16}
\int dy \left[
\left(  \frac {(z-y)^2}{{z}^{3}} - 2 N_f \right) \Ffmt
\right.
\nn
\\
&& \left. -\Nf^{2} \, \left( \Fft -\Ffmt \right) + 2 (\Delta_0
\beta ) \Nf (1-\Nf)\left( 1- 2 \dtfob\right) \right] , \nn
\\
\gamma^{{\rm Db}}_{1,0} & = &
-\frac{ \rho \beta}{16}
\int dy \left[
 \left(   \frac {(z+y)^2}{{z}^{3}} - 2 N_f \right) \Ffmt
\right.
\nnel
&&
\left.
- \Nf^{2} \, \left( \Fft -\Ffmt \right)
+ 2 (\Delta_0 \beta ) \Nf (1-\Nf)\left( 1- 2 \dtfob\right)
\right]
,
\nn
\\
\gamma^{{\rm ND}}_{1,0} & = & - \frac{\rho\,\beta}{16} \int dy \left[
\left(- \frac{1}{ z^{3}} + 2 N_f \right) \Ffmt + \Nf^{2} \left(
\Fft-\Ffmt \right ) \right. \nnel && \left. +2 (\Delta_0 \beta)
\Nf (1-\Nf)  \left(1- 2 \dtfob \right) \right] .
\label{gammaone}
\end{eqnarray}
Notice that the only difference
between these two diagonal contributions $\gamma^{{\rm Da}}_{1,0}$ and 
$\gamma^{{\rm Db}}_{1,0}$ are the terms odd in $y$
(i.e.\ $\epsk$). One should also note that for $t>t_i$, the
$\gamma_{1,0}$ terms {\em contain both real and imaginary
components.}  In the equations of motion, which follow directly
from (\ref{ane}), these correspond to oscillatory and dissipative
terms respectively. The second order terms may be found in
appendix \ref{crap}.

The above results may be rewritten in terms of the real Higgs
$\phi_h$ and Goldstone fields $\phi_g$ where
$$
\phi =
\frac{1}{\sqrt{2}}(\phi_h + i \phi_g)  .
$$
One obtains
\begin{eqnarray}
\Seffbar &:=&  - i {\rm tr} \intdfx \ln \matr{G}^{-1}(x,x) -
\frac{1}{g} \intdfx \Delta_0^2
 \nn
 \\
& +&   \intdfx
 \left(
    \phi_h \gamma^{H}  \phi_h
+   \phi_g \gamma^{G}  \phi_g
+ i \phi_h \gamma^{HG} \phi_g
- i \phi_g \gamma^{HG} \phi_h
\right).
\nn
\end{eqnarray}
The coefficients $\gamma^h$ and $\gamma^g$ are defined in the derivative
expansion
as in \ref{gammaexp} and their explicit forms are
\begin{eqnarray}
\gamma^H_{0,0} & = &  2\gamma^{Da}_{0,0}
\; \;  =
 \; \; \frac{ \rho }{4} \int dy \left[
 \,{\frac {1}{{z}^{3}}}
- 2\Nf
- 2(\Delta_0 \beta) N_f (1-N_f)
\right]
\nn
\\
\gamma^G_{0,0} & = & 0
\nn
\\
\gamma^{H}_{1,0}
& = &
- \frac{ \rho \beta}{8} \int dy
\left[
\frac {y^2}{z} \Ffmt
+ 2 (\Delta_0 \beta ) \Nf (1-\Nf) \dtfob
\right]
\nnel
\gamma^{H}_{0,1} & = & 0
\nn
\\
\gamma^{G}_{0,0} & = & \gamma^{G}_{0,1} = 0
\nn
\\
\gamma^{G}_{1,0}
& = & -\frac{\rho \beta}{8}
\int dy
\left[
\left( \frac {1}{z} - 2 \Nf  \right) \Ffmt
- \Nf^{2} \left( \Fft-\Ffmt \right )
\right]
\nn
\\
\gamma^{HG}_{0,0} & = & \gamma^{HG}_{0,1} = 0
\nn
\\
\gamma^{HG}_{1,0} & = &  \frac{\rho \beta}{8} \int dy \frac
{y}{z^2} \Ffmt \label{strange}.
\end{eqnarray}

% ******************************************************************
\section{Discussion and Conclusions}
\label{secconc}

Several comments can be made about these results.  First, for any
static field configurations $E''=0$, and hence  $\exp \{ \beta E''
\}=1$ as assumed in all the previous calculations (see the
discussion in section \ref{secde}). Thus for static configurations
we recover the static limit of
retarded equilibrium Green function results used by
others, so that our effective action is identical to standard
calculations in the static limit.  In
particular, our lowest order term in the derivative expansion, the
effective potential, is identical to the one which would be calculated
using pure equilibrium methods.  The differences are in the
non-static terms:  we have a unique derivative expansion which
does {\em not} depend on the order in which the spatial and
temporal derivative expansions are made. Thus only by using our
expansion can one justify the use of the traditional effective
potential as the potential for some effective theory, since here
we have a
fully defined and controlled expansion. Following on
from this, since the physical masses for the bosonic field
$\Delta$ are encoded by the second derivative of the effective
potential, we expect to reproduce Goldstone's theorem.
We indeed find that the mass term for the Goldstone mode,
$\gamma^G_{0,0}$, is zero.

As discussed in section \ref{secde}, our calculation takes into
account the fact that time translation invariance is broken by the
non-equilibrium derivative expansion.  This is reflected in the
final effective action obtained in (\ref{ane}) where the
coefficients depend explicitly on the initial time $t_i$ at which
the system was perturbed.

Though we believe that the methods presented here are more
consistent than previous methods used to calculate the derivative
expansion, there are still a number of confusing aspects in this
work.  One of these (which we believe also exists in the previous
calculations, though it is not explicitly commented upon there) is
the question of the difference between $\Sefft[{\rm Da}]$ and
$\Sefft[{\rm Db}]$ of \tref{Da} and \tref{Db}.  We believe to have
argued correctly in
section \ref{seccalc} that the change of variables $t
\leftrightarrow t'$ and $\vecp \rightarrow -\vecp$ should {\em
not} be carried out before the derivative expansion.  However,
this in turn is responsible for the
coupling between the Higgs and Goldstone modes in \tref{strange}.
Another interesting aspect of our results is that apart from
breaking time-translation invariance (which we {\em do} believe to
be correct), observe that another symmetry has been broken:- since
$\left(\gamma_{1,0}^{\rm Da} \right)^* \neq \gamma_{1,0}^{\rm Db}$
then to this order in the derivative expansion $\Seffbar$ is not
invariant under $\phi \leftrightarrow \phi^*$.  However, one fully
expects this to be broken if there is a non-zero charge in the system.
This corresponds to having an imaginary initial time derivative for
the scalar field $\phi$.  Finally, note that the first terms of
$\gamma_{1,0}^{\rm Da} $ and $ \gamma_{1,0}^{\rm Db}$ are actually
divergent at $t=t_i$.  This infinity is exactly the one which
arises in (\ref{Zbar}) if one were to set $t=t_i$ and $\vecp=0$.
It should be noted that this divergence is not present in the
usual calculations.  As commented on just after equation
(\ref{shorthand}) it stems from the fact that at $t=t_i$ our
expressions differ from similar terms in \cite{St,AL} by the factor
of $e^{\beta E''} \neq 1$.  In this non-relativistic case such infinities
are not too worrying one always has a UV cutoff in mind, but the
same is not true in relativistic cases where they would have to be
renormalised in some way \cite{TSEea}.

To conclude, we have argued that the traditional approach to
producing derivative expansions of thermal effective actions are
inconsistent since they try to use equilibrium fields in
constructing the derivative expansion.  We have argued that this
is technically not correct, and that it is incompatible with the
physical problem being described---that of slowly varying fields.
We have shown that letting the fields be out of equilibrium leads
to a well behaved derivative expansion. It means that the Green
functions at the heart of our calculations, our $\calB$'s, are not
the retarded thermal Green functions normally encountered in such
work.  Indeed, as linear response analysis shows \cite{lB}, those
are relevant to a very different type of physical problem where
the system is hit with a sudden change rather than responding to a slow
variation.\tnote{e.o.m.? Infinities?}  Finally, we have used
these ideas to calculate the effective action for a BCS
superconductor.

\section*{Acknowledgements}

We thank I.Aitchison, I.Lawrie, R.Rivers and A.Schakel for useful
discussions, and also G.Vitiello and Salerno University for
hospitality.
D.A.S.\ is supported by P.P.A.R.C.\ of the UK through a
research fellowship and is a member of Girton College, Cambridge.
This work was supported in part by the
\href{http://www.esf.org/}{European Science Foundation},
with additional support from
the European Commission through their Socrates programme.

% ****************************************************************
% Appendices

\renewcommand{\thesection}{\Alph{section}}
\setcounter{section}{0}

\section{Calculation of the fermionic propagator}\label{appdp}

Calculation of the fermion propagator in $(t,\veck)$
variables proceeds as follows.  Throughout it is vital to keep
track of the delta functions and boundary conditions so that
the solution is valid for
{\em any} path-ordered formalism of
thermal field theory.

As in section \ref{seccalc}, define the matrix $\matG^{-1}(x)$ through
\begin{equation}
\matG^{-1}(x) \delta^4(x-y) := \matG^{-1}(x,y)
\end{equation}
so that in $(t,\bfk)$ space
\begin{eqnarray}
\matG^{-1}(t,\bfk) &=&
\left(
\begin{array}{cc}  i \paa_t -
\frac{\bfk^2}{2m} + \mu & \Delta_0 \\
\Delta_0^{} &   i \paa_t +
\frac{\bfk^2}{2m} - \mu
\end{array} \right)
=
\left(
\begin{array}{cc}  i \paa_t -
\epsk & \Delta_0 \\
\Delta_0^{} &   i \paa_t +
\epsk
\end{array} \right)
\nonumber
\\
& = & i \paa_t \matone + \matr{A}_{\bfk}.
\label{somatx}
\end{eqnarray}
Here $ \epsk = \frac{\bfk^2}{2m} - \mu$,
\begin{equation}
\matr{A}_{\bfk}
= \left(
\begin{array}{cc}
    - \epsk & \Delta_0 \\
\Delta_0^{} &    + \epsk
\end{array} \right)
= \Ek
\left(
\begin{array}{cc}
- \cos ( 2 \theta_\veck ) &  \sin ( 2 \theta_\veck )  \\
  \sin ( 2 \theta_\veck ) &  \cos ( 2 \theta_\veck )  \\
\end{array} \right),
\label{Adefx}
\end{equation}
and $\Ek$ is the dispersion relationship $\Ek^2 =  |\Delta_0|^2 +
\epsilon^2_\bfk $.  The angle $\theta_\veck$ is defined through
\begin{equation}
\cos ( 2 \theta_\veck ) = \frac{\epsk}{\Ek} , \; \; \; \sin ( 2
\theta_\veck ) = \frac{\Delta_0}{\Ek} = - 2 | u_+(\veck) | |
u_-(\veck)  |
\end{equation}
where, as in section \ref{seccalc},
\begin{equation}
u_\pm(\veck)
= \left( \half \left[ 1 \pm \frac{\epsk}{\Ek} \right] \right)^\half
\end{equation}
so that $\cos ( \theta_\veck ) = u_+(\veck)$ and $- \sin (
\theta_\veck ) = u_-(\veck)$.\tnote{In the operator method $u_+$
and $u_-$ are the coefficients of the Bogoliubov transformation
which diagonalises the Lagrangian, and they are usually denoted by
$u$ and $v$ respectively.}

Now, from equation (\ref{propG}),
\begin{eqnarray}
 \matG^{-1}(x) \matG(x,z) &=& \delta^{4}(x-z) \matone
 \nonumber
 \\
 \Longrightarrow \; \; \; \; \; \; \; \; \; \; &&
\nonumber
\\
(i \paa_t \matone + \matr{A}_{\bfk}) \;
\matG(t,t',{\bfk}) & =& \delta(t-t') \matone .
\label{importantx}
\end{eqnarray}
The diagonalised form of $\matr{A}_\veck$ is
\begin{equation}
\matr{A}_\veck = \matr{B}_\veck \cdot \matr{\Lambda}_\veck
 \cdot \matr{B}_\veck^{-1}
\end{equation}
where $\matr{\Lambda}_\veck$ contains the eigen-values of
$\matr{A}_\veck$;
\begin{equation}
\matr{\Lambda}_\veck = \left(
\begin{array}{cc}
\Ek & 0 \\
0 &   -\Ek
\end{array} \right),
\label{Lambdadefx}
\end{equation}
and $\matr{B}_\veck$ its eigen-vectors;
\begin{equation}
\matr{B}_\veck
=
\left(
\begin{array}{cc}  -
\sin ( \theta_\veck ) &    \cos ( \theta_\veck )  \\
\cos ( \theta_\veck ) &  - \sin ( \theta_\veck )  \\
\end{array} \right)
 =
\left(
\begin{array}{cc} u_-(\veck) &  u_+(\veck)   \\
u_+(\veck) & u_-(\veck)  \\
\end{array} \right)  .
\label{Bdefx}
\end{equation}
Now equation (\ref{importantx}) is equivalent to
\begin{equation}
(i \paa_t \matr{B}_\veck^{-1} + \matr{B}_\veck^{-1} \matr{A}_\veck) \;
\matG(t,t',\veck)\matr{B}_\veck
= \delta(t-t')
\matone
\end{equation}
so that on defining the matrix
\begin{equation}
\matr{F}(t,t',\veck) = \matr{B}_\veck^{-1} \matG(t,t',\veck) \matr{B}_\veck
\label{Fdefx}
\end{equation}
one obtains
\begin{equation}
(i \paa_t \matone +  \matr{\Lambda}_\veck) \;
\matr{F}(t,t',\veck)
= \delta(t-t')
\matone.
\label{Greenx}
\end{equation}
Thus
$\matr{F}_\veck$ satisfies the usual Feynman equation and is given by
\begin{equation}
\matr{F}(t,t',\veck) =  \left(
\begin{array}{cc}
F_+(t,t',\veck) & 0 \\
0 &   F_-(t,t',\veck)
\end{array} \right)
\label{Fanswerx}
\end{equation}
where in the Mills representation \cite{Mills}
\begin{eqnarray}
F_\pm (t,t',\veck) &=& -i \int
\frac{dk_0}{2 \pi} e^{-i k_0 (t-t') } \left( \theta_C(t,t') -
N(k_0) \right) \rho_\pm(k_0,\bfk) \nn
\label{freeFx}
\end{eqnarray}
so that $F_- (t,t',\veck) =  F_+ (t',t,\veck)$.
Here $N(k_0)$ is the Fermi-Dirac distribution:
\begin{equation}
N(k_0) = \frac{1}{e^{\beta k_0} + 1},
\end{equation}
the $\theta_{C}$ is the usual $\theta$ function for the contour $C$
\cite{LvW,lB},
and the spectral functions $\rho_\pm$ are given by
\begin{equation}
\rho_\pm(k_0,\bfk) = 2\pi \delta(k_0 \pm \Ek)   .
\label{rhox}
\end{equation}
Finally, these results may be used to compute the explicit form of
$\matG(t,t',\bfk)$:  inverting equation  (\ref{Fdefx}), using (\ref{Bdefx})
and
(\ref{Fanswerx}) gives, after some manipulation,
\begin{equation}
\matG(t,t',\bfk) =  \left(
\begin{array}{cc}
u^2_+ F_-(t,t',\bfk) + u^2_- F_+(t,t',\bfk) & u_+
u_- (F_+(t,t',\bfk)
- F_-(t,t',\bfk) ) \\
u_+ u_- (F_+(t,t',\bfk)- F_-(t,t',\bfk))  &
u^2_+ F_+(t,t',\bfk) +
u^2_- F_-(t,t',\bfk)
\end{array} \right)
\label{AnswerGapp}
\end{equation}
where the $\veck$ dependence of $u_\pm(\veck)$ has been omitted but is
understood.

% *****************************************************************

\section{Second Order Terms}\label{crap}

Here we write out the results for the second order terms.

\begin{eqnarray}
\gamma^{{\rm Da}}_{2,0} &=& -\frac{\rho}{16 \Delta_0^2} 
\int dy \; 
\left\{
\frac{z^2+y^2}{z^5}
+ \frac{(z-y)^2}{z^4}
(\beta \Delta_0) \Ffmt
[(\beta \Delta_0-2 \Delta_0 \dtf) z-1]  \right.
\nnel
&&
\tsespace
\left.
 + \Nf [-2+(4/3) (\Delta_0 \beta)^3]
 - \Nf^2 \Fft (\beta \Delta_0) [1+\beta \Delta_0- 2 \Delta_0 \dtf ]
\right.
\nnel
&& \tsespace  \left.
 - 4 \Nf (1-\Nf) (\beta \Delta_0) (\Delta_0 \dtf) [\beta \Delta_0 - \Delta_0
 \dtf ]
\right.
\nnel
&& \tsespace  \left.
 + (2-\Nf) \Nf \Ffmt (\beta \Delta_0) [1-(\beta \Delta_0- 2 \Delta_0
 \dtf) ]
\right\}
\nn
\\
\gamma^{{\rm Db}}_{2,0} &=& \left. \gamma^{{\rm Da}}_{2,0}\right|_{y
\rightarrow - y}
\nn
\\
&=& -\frac{\rho}{16 \Delta_0^2} \int dy \; 
\left\{ \frac{z^2+y^2}{z^5} + \frac{(z+y)^2}{z^4}
  (\beta \Delta_0) \Ffmt
  [(\beta \Delta_0-2 \Delta_0 \dtf) z-1]
\right.
\nnel
&& \tsespace  \left.
 + \Nf [ -2+(4/3) (\Delta_0 \beta)^3 ]
 - \Nf^2 \Fft (\beta \Delta_0)
 [ 1+\beta \Delta_0- 2 \Delta_0 \dtf]
\right.
\nnel
&& \tsespace  \left.
 - 4 \Nf (1-\Nf) (\beta \Delta_0) (\Delta_0 \dtf) [\beta \Delta_0 - \Delta_0
 \dtf]
\right.
\nnel
&& \tsespace  \left.
 +  \Nf (2-\Nf) \Ffmt (\beta \Delta_0) [1-(\beta \Delta_0- 2 \Delta_0
 \dtf)]
\right\}
\nn
\\
\gamma^{{\rm ND}}_{2,0}
&=&
-\frac{\rho}{16 \Delta_0^2} \int dy \; 
\left\{
-\frac{1}{z^5} + \frac{1}{z^4}   (\beta \Delta_0) \Ffmt
[1-z (-2 \Delta_0 \dtf+\beta \Delta_0)] \right.
\nnel
&& \tsespace  \left. +  \Nf [2+(4/3) (\Delta_0 \beta)^3]
 + \Nf^2 \Fft (\beta \Delta_0) [1+\beta \Delta_0- 2 \Delta_0 \dtf ]
\right.
\nnel
&& \tsespace  \left.
 - 4 \Nf (1-\Nf) (\beta \Delta_0) (\Delta_0 \dtf) [\beta \Delta_0 - \Delta_0
 \dtf]
\right.
\nnel
&& \tsespace  \left.
 - \Nf (2-\Nf) \Ffmt (\beta \Delta_0) [1-(\beta \Delta_0- 2 \Delta_0
 \dtf)]
\right\}
\nn
\\
\gamma^{{\rm Da}}_{0,2} &=& \gamma^{{\rm Db}}_{0,2}
\nn
\\
&=& \frac{\rho \mu}{48 m \Delta_0^2} \int dy \; 
\left\{ \frac{1}{z^7} \left[
 (-16 z^2+10) +  \frac{\Delta_0}{\mu} y (-6 z^4-23 z^2+20) \right. \right.
\nn
\\
&& \tsespace 
\left.  +  \left(\frac{\Delta_0}{\mu}\right)^2 (-6 z^6-z^4+17 z^2-10)
 \right]
\nnel
&& \tsespace
\left.+
 4 \Nf  \left[ 3+3 (\beta \Delta_0) -(\beta \Delta_0)^2 \right]
 +12 \Nf^2 (\beta \Delta_0) \left(-1 + (\beta \Delta_0) \right)
  \right\}
\nn
\\
\gamma^{{\rm ND}}_{0,2} &=&
\frac{\rho \mu}{48 m \Delta_0^2} \int dy \; 
\left\{
\frac{1}{z^7}\left[
(-4 z^2+10) +  \frac{\Delta_0}{\mu} y (z^2+20)
 +  5 \left(\frac{\Delta_0}{\mu}\right)^2 (z^4+z^2-2)\right]
\right.
\nnel
&& \tsespace
\left.
 -4 \Nf  \left[ 3+3 (\beta \Delta_0) +(\beta \Delta_0)^2 \right]
 +12 \Nf^2 (\beta \Delta_0) \left(1 + (\beta \Delta_0) \right)
\right\}.
\nn
\end{eqnarray}

% ************************************************************************
%\newpage
\typeout{--- No new page for bibliography ---}

\end{document}